\documentclass[3p,twocolumn]{elsarticle}

\usepackage{amssymb}

\journal{arXiv}

\usepackage{comment}
\usepackage{subfig,pgfplots,epsf,graphicx,color}
\usepackage{amsmath,amsfonts,amsthm,multirow}

\usepackage{rotating}
\usepackage{tabularx}
\usepackage{float}

\usepackage{pdfpages}
\usepackage{afterpage}
\usepackage{wrapfig}
\usepackage{microtype}
\usepackage{fullpage}
\usepackage{bm} 
\usepackage{caption} 

\usepackage[colorlinks=true]{hyperref}
\hypersetup{
  colorlinks  = true, 
  urlcolor    = blue, 
  linkcolor   = blue, 
  citecolor   = blue, 
}

\DeclareGraphicsExtensions{.gif,.png,.jpg,.pdf,.tiff,.eps}

 \usepackage{lineno,hyperref}
\modulolinenumbers[5]

\usepackage{tikz}
\usepackage{pgfplots}
\usepackage{pgfmath}
\usepackage[ruled,vlined,linesnumbered]{algorithm2e}

\newcommand{\dt}{\dftl{t}}

\newcommand{\cJ}{\mathcal{J}}

\newcommand{\be}{\begin{equation}}
\newcommand{\ee}{\end{equation}}
\newcommand{\bes}{\begin{equation*}}
\newcommand{\ees}{\end{equation*}}
\newcommand{\bse}{\begin{subequations}}
\newcommand{\ese}{\end{subequations}}

\newcommand{\uu}{\mathbf{u}}

\newcommand{\ux}{u^1}

\def\dO{\partial \Omega}

\def\ee{{\hat {\underline e}}}

\def\dt{ \Delta t }

\def\cJ{{\cal J}}

\def\scriptO{{{\it O}\kern -.42em {\it `}\kern + .20em}}
\def\RR{{{\rm l}\kern - .15em {\rm R} }}
\def\PP{{{\rm l}\kern - .15em {\rm P} }}
\def\L2{{{\sf L}^2}}
\def\H1{{{\sf H}^1}}

\def\PN2{{\PP_{N}-\PP_{N-2}}}

\def\complex{{{\rm C} \kern - .53em {\rm l} \kern + .38em}}

\def\a1{{ | \lambda_{\min} |}}

\def\l1{{   \lambda_{\min}  }}

\def\btu{{\tilde {\bf u}}}

\def\bu0{{\underline {\bf 0}}}

\def\br{{\bf r}}
\def\bu{{\bf u}}
\def\bv{{\bf v}}
\def\bx{{\bf x}}

\def\bD{{\bf D}}

\def\Ab{{\bar A}}

\def\ih{{\hat \imath}}
\def\Ih{{\hat I}}
\def\jh{{\hat \jmath}}
\def\eh{{\hat e}}

\def\kh{{\hat k}}

\def\Dh{{\hat D}}

\def\Oh{{\hat \Omega}}

\def\ub{{\underline b}}

\def\ug{{\underline g}}

\def\up{{\underline p}}

\def\uu{{\underline u}}

\def\uv{{\underline v}}
\def\uw{{\underline w}}
\def\ux{{\underline x}}

\def\u0{{\underline 0}}
\def\n12{{n_{\frac{1}{2}}}}
\def\t12{{t_{\frac 1 2}}}
\def\n12{{n_{0.8}}}
\def\t12{{t_{0.8}}}
\def\bxi{{\boldsymbol \xi}}

\newcommand{\pp}[2]{\frac{\partial #1}{\partial #2} }
\newcommand{\dd}[2]{\frac{d #1}{d #2} }

\begin{document}

\begin{frontmatter}




 \title{NekRS, a GPU-Accelerated Spectral Element Navier--Stokes Solver}

 
\author[2,3,1]{Paul Fischer}
\author[1]{Stefan Kerkemeier}
\author[1]{Misun Min\corref{mycorrespondingauthor}}\ead{mmin@mcs.anl.gov}
\author[1]{Yu-Hsiang Lan}
\author[2]{Malachi Phillips}
\author[2]{Thilina Rathnayake}
\author[4,1]{Elia Merzari}
\author[5,1]{Ananias Tomboulides}
\author[6]{Ali Karakus}
\author[7]{Noel Chalmers}
\author[8]{Tim Warburton}

\address[1]{Mathematics and Computer Science, Argonne National Laboratory, Lemont, IL 60439}
\address[2]{Department of Computer Science, University of Illinois at Urbana-Champaign, Urbana, IL 61801}
\address[3]{Department of Mechanical Science and Engineering, University of Illinois at Urbana-Champaign, Urbana, IL 61801}
\address[4]{Department of Nuclear Engineering, Penn State, PA 16802}
\address[5]{Department of Mechanical Engineering, Aristotle University of Thessaloniki, Greece 54124}
\address[6]{Mechanical Engineering Department, Middle East Technical University, 06800, Ankara, Turkey}
\address[7]{AMD Research, Advanced Micro Devices Inc., Austin, TX 78735}
\address[8]{Department of Mathematics, Virginia Tech, Blacksburg, VA 24061}
\cortext[mycorrespondingauthor]{Corresponding author}



 \begin{abstract}
 
The development of NekRS, a GPU-oriented thermal-fluids simulation code based
on the spectral element method (SEM) is described.  For performance portability,
the code is based on the open concurrent compute abstraction  and
leverages scalable developments in the SEM code Nek5000 and 
in libParanumal, which is a library of high-performance kernels 
for high-order discretizations and PDE-based miniapps.   Critical performance
sections of the Navier--Stokes time advancement are addressed.  Performance
results on several platforms are presented, including scaling to 27,648
V100s on OLCF Summit, for calculations of up to 60B gridpoints.


 \end{abstract}



 \begin{keyword}
  NekRS \sep
  Nek5000 \sep
  libParanumal \sep
  OCCA \sep
  GPU \sep
  Scalability \sep
  Performance \sep
  Spectral Element Method \sep
  Incompressible Navier--Stokes \sep 
  Exascale Applications
 \end{keyword}

\end{frontmatter}



%

\section{Introduction}

A fundamental challenge in fluid mechanics and heat transfer is to accurately
simulate physical interactions over a large range of spatial and temporal
scales.  Such simulations can involve billions of degrees of freedom evolved
over hundreds of thousands of timesteps.  Simulation campaigns for these
problems can require weeks or months of wall-clock time on the world’s fastest
supercomputers.  One of the principal objectives of high-performance computing
(HPC) is to reduce these runtimes to manageable levels.

We are interested in modeling turbulent flows using either direct
numerical simulation (DNS) to capture all scales of motions, large eddy
simulation (LES) to capture the modes that dominate momentum and thermal
transport, or Reynolds-averaged Navier--Stokes (RANS) formulations that emulate
both small- and large-scale transport with closure models.  
Applications include reactor thermal hydraulics, internal combustion
engines, ocean and atmospheric flows, vascular flows, astrophysical problems,
and basic turbulence questions for theory and model development.  Simulations
in these areas present significant challenges with respect to scale resolution,
multiphysics, and complex computational domains. In many cases, experimental
data are expensive or impossible to obtain, making simulation on leadership
computing platforms critical to informed analysis.  

With current exascale computing programs in the U.S. and elsewhere developing
GPU-based HPC platforms it is imperative to exploit the performance potential
of these powerful node architectures.  In this paper, we describe the
development of a new GPU-oriented open-source code for thermal-fluid analysis,
NekRS, which has emerged out of two HPC software projects. {\em Nek5000} \cite{nek}
was one of the first production-level single-program multiple-data (SPMD) codes
deployed on distributed-memory parallel computers \cite{dd88}.  It has
demonstrated scalability to leading-edge platforms through the SPMD era
\cite{tufo99a,fischer08a} and readily scales to millions of MPI ranks
\cite{fischer15}.  
   Early GPU efforts for Nek5000 commenced with OpenACC
ports \cite{gong2015} and \cite{min2015a} (for NekCEM).
{\em libParanumal} \cite{warburton2019,warburton2019b} 
is a self-contained high-order finite
element library that uses highly optimized kernels based on the portable Open
Concurrent Compute Abstraction (OCCA) \cite{medina2015okl,occa}.  It
includes sublibraries for dense linear algebra, Krylov solvers, parallel mesh
handling and polynomial approximation, $p$-type and algebraic multigrid, time
stepping, gather-scatter operations and halo exchanges, and core miscellaneous
operations. 
The libParanumal sublibraries support meshes consisting of
triangles, quadrilaterals, tetrahedra, or hexes. The libParanumal project also
includes mini-apps providing GPU accelerated solvers for a wide variety
of transport-dominated physics applications.  A significant feature of 
the libParanumal kernels is that, in the majority of cases, they are tuned
to meet the roofline performance limits.  For example, FP64 performance
in excess of 1 TFLOPS is realized for local SEM matrix-vector product (matvec)
kernels on the NVIDIA V100 \cite{warburton2019,fischer20a}.
Each solver supports multi-GPU simulation via Nek5000's {\em gslib} for
efficient MPI-based gather-scatter operations and halo exchanges \cite{gslib}.

In the present work, we describe a new code, NekRS, which is written in
C++/OCCA. The performant kernels in NekRS started as an early fork from
libParanumal and were tailored and expanded to meet the specific requirements
of large-scale turbulent flow applications in complex domains.  NekRS provides
access to the standard Nek5000 interface and features (e.g., conjugate heat
transfer), which allows users to leverage existing application-specific source
code and data files on GPU-based platforms.

The remainder of this article is organized as follows.
In Section 2 we provide relevant details of the governing
       equations and spectral element discretization.
In Section 3 we describe the parallel GPU development, 
   including parallel communication and partitioning 
   strategies at exascale and present  an illustration of high-performance
   kernels for GPU-based nodes.
In Section 4 we provide extensive performance studies at scale, 
  including weak- and strong-scale studies on all of Summit.
We conclude with remarks and discussion in Section 5.

\section{Formulations}       

\def\bw{{\bf w}}

We simulate thermal transport governed by the incompressible Navier--Stokes (NS) 
and energy equations,
\index{Navier-Stokes} 
\begin{eqnarray} \label{eq:ns}
\pp{\bu}{t} + \bu\cdot\nabla \bu &=& -\nabla p + \frac{1}{Re} \nabla^2 \bu,
\\ \label{eq:inc}
\qquad \nabla \cdot \bu &=& 0, 
\\ \label{eq:energy}
\pp{T}{t} + \bu\cdot\nabla T   &=&             \frac{1}{Pe} \nabla^2 T,
\end{eqnarray} 
subject to appropriate velocity ($\bu$), pressure ($p$),
and temperature ($T$) initial conditions in $\Omega$ and
boundary conditions on $\dO$.
For typical applications, the Reynolds ($Re$) and Peclet ($Pe$) numbers
are large, implying that the flows are advection dominated.
For high $Re$, in fact, the flows are fully turbulent, implying
a need for highly accurate numerical discretizations in order to 
avoid numerical dispersion and dissipation \citep{kreiss72}.
(We note that NekRS currently supports conjugate heat transfer 
where $T$ may be defined on a domain that is larger than $\Omega$.
In what follows, however, we omit further discussion of the energy equation
(\ref{eq:energy}).)

\subsection{BDF Time Discretization}
We begin with a backward difference (BDF$k$) approximation to $\pp{\bu}{t}$
to derive an implicit Stokes substep for velocity and pressure at
time level $t^n$,
\begin{eqnarray} \label{eq:ns_1a} 
\hspace*{-.3in} &&
\frac{\beta_0}{\dt} \bu^{n}=\frac{1}{\dt} \bu^*
- \nabla p^n + \frac{1}{Re}\nabla^2 \bu^n, 
\\ \label{eq:ns_1b}  &&
\qquad \nabla \cdot \bu^n = 0, 
\end{eqnarray} 
where 
$\bu^*$ accounts for quantities known from prior substeps and is computed in
one of two ways.  The standard (Courant- or CFL-limited) formulation is
BDF$k$/EXT$k$,
\begin{eqnarray} \label{eq:stdadv}
\hspace*{-.5in} &&
  \bu^* := - \sum_{j=1}^k \left(
   \beta_{j}\bu^{n-j}+ \dt\, \alpha_{j}\bu^{n-j}\cdot\nabla\bu^{n-j}\right),
\end{eqnarray} 
where the $\beta_j$s are the $k$th-order BDF coefficients and 
the $\alpha_j$s are $k$th-order extrapolation coefficients.
An alternative formulation that avoids the CFL constraint on stepsize
$\dt$ is the semi-Lagrangian approach with
\begin{eqnarray} \label{eq:oifs}
   \bu^* &:=& -  \sum_{j=1}^k \beta_{j}\btu^{n-j}.
\end{eqnarray} 
Here, each $\btu^{n-j}(\bx)$ represents the value of $\bu^{n-j}(\bx^*),$
where $\bx^*$ is the foot of the characteristic that would be found by 
integrating the velocity field backward in time over $[t^n,t^{n-1}]$.
In practice, the off-grid interpolation required for direct evaluation 
of $\bu^{n-j}(\bx^*)$ can be avoided by solving a hyperbolic advective
subproblem on $[t^{n-j},t^n]$,
\begin{eqnarray} \label{eq:hyp}
\pp{\bw}{t} + \bu \cdot \nabla \bw = 0,
\end{eqnarray}
with initial condition $\bw(\bx,t^{n-j})=\bu^{n-j}$ \citep{maparo90,saumil18}.
We typically use the third-order ($k=3$) formulation for (\ref{eq:stdadv}) with
a Courant number of CFL=0.5. For (\ref{eq:oifs}), CFL=2--4 is most common, but
we typically use only second-order in time ($k=2$) because of the relative
expense of the hyperbolic substeps (\ref{eq:hyp}), which are fully dealiased
\cite{johan13}.

\subsection{Implicit Stokes Solve}
The unsteady linear Stokes problem (\ref{eq:ns_1a})--(\ref{eq:ns_1b}) is
further decoupled via a fractional step method that treats the divergence-free
and viscous terms as separate subproblems.  
A pressure-Poisson problem derives from taking the divergence of (\ref{eq:ns_1a}),
\begin{eqnarray} \label{eq:ns_2c}
\hspace*{-.2in}
- \nabla \cdot (\nabla p^n) = -\frac{\nabla \cdot \bu^*}{\Delta t}
+ \frac{1}{Re} \nabla \cdot  (\nabla \times \omega),
\end{eqnarray} 
where $\omega = \sum_{j=1}^k \alpha_j \nabla \times \bu^{n-j}$ is the 
extrapolated vorticity, which serves to control divergence errors at the
boundaries.  Additional details on the boundary conditions for
(\ref{eq:ns_2c}) can be found in 
\citep{orsisrde86,tomboulides89,tomb97,guermond06}. 

The final substep requires the solution of 
\begin{eqnarray} \label{eq:ns_2d}
- \frac{1}{Re} \nabla^2 \bu^n  + \frac{\beta_0}{\Delta t}\bu^n =
  \frac{\bu^{**}}{\Delta t}, 
\end{eqnarray} 
where $\bu^{**}$ is the divergence-free velocity 
\begin{eqnarray} \label{eq:ns_2e}
   \bu^{**} = \bu^* - \Delta t \nabla p^{n}.  
\end{eqnarray} 

The advantage of (\ref{eq:ns_1a})--(\ref{eq:ns_2e}) is that it decouples the NS
equations into independent substeps, each of which can be efficiently treated
by techniques tailored to the governing physics: hyperbolic substeps for
advection, diagonally preconditioned conjugate gradient (PCG) iteration for the
viscous Helmholtz problems, and multilevel PCG or GMRES for the pressure
solve.  Because it governs the fastest modes (i.e., the acoustic
modes, which are infinitely fast in the incompressible model), the 
pressure-Poisson problem is intrinsically the stiffest substep.  Isolating it from the
other governing operators results in a fast algorithm because there is no need to
evaluate viscous or advection operators with each iteration, which would be 
required of a fully implicit approach.

\subsection{Spectral Element Discretization}

To develop an efficient spatial discretization,
we employ high-order spectral elements (SEs) \citep{pat84} in
which the solution, data, and test functions  are represented as {\em locally
structured} $N$th-order tensor product polynomials on a set of $E$ {\em globally
unstructured} curvilinear hexahedral brick elements.   The approach yields two
principal benefits.  First, for smooth functions such as solutions to the
incompressible NS equations, high-order polynomial expansions yield exponential
convergence with approximation order, implying a significant reduction in the
number of unknowns ($n \approx EN^3$) required to reach engineering tolerances. 
Second, the locally structured forms permit local lexicographical ordering with
minimal indirect addressing and, crucially, the use of tensor-product sum
factorization to yield low $O(n)$ storage costs and $O(nN)$ work complexities 
\citep{sao80}.  As we demonstrate, the leading order $O(nN)$ work terms can be
cast as small dense matrix-matrix products (tensor contractions) with favorable
$O(N)$ work-to-storage ratios (computational intensity)  \citep{dfm02}.

The equations for the SE basis coefficients are derived from a weighted
residual formulation for each subproblem.  For example,
(\ref{eq:stdadv}) and (\ref{eq:ns_2d})--(\ref{eq:ns_2e}) become
{\em Find $\bu^n,p^n \in X_b^N(\Omega)$ $\times \, X^N(\Omega)$ such that}
{\em for all $\bv,q \in X_0^N$ $\times \, X^N$},
\hspace*{-.3in}
\begin{eqnarray} \nonumber
\hspace*{-.3in} 
&&
\hspace*{-.6in} 
(\bv,\bu^*)  = - \sum_{j=1}^k \left( \beta_{j}
                  (\bv, \bu^{n-j}) +  \right. \\ \label{eq:ns_wrt0} &&
                  \left. \dt\, \alpha_{j}
                  (\bv, \bu^{n-j}\!\cdot\!\nabla\bu^{n-j})\right),
\\[1.0ex] \label{eq:ns_wrt1}  &&
\hspace*{-.6in}
(\nabla q,\nabla p^n)= - \frac{1}{\dt} (q,\nabla \cdot \bu^*) 
                      \! -\! \frac{1}{Re} (\nabla q,\nabla \times \omega)
\\[1.3ex] \label{eq:ns_wrt2} &&
\hspace*{-.6in}
\frac{1}{Re}
(\nabla \bv,\nabla \bu^n)+\frac{\beta_0}{\dt}(\bv,\bu^n) = \frac{1}{\dt}
(\bv,\bu^{**}).
\end{eqnarray} 
Here, $(\bv,\bu)=\int_\Omega~\bv~\cdot~\bu~dV$ is the $L^2$ inner product
on $\Omega$;
$X^N_b$ is the subset of $X^N$ satisfying the Dirichlet conditions on $\dO$;
$X^N_0$ is the subset of $X^N$ satisfying homogeneous Dirichlet conditions on
$\dO$; and $X^N \subset H^1$ is the set of continuous $N$th-order
spectral element basis functions described in \citep{dfm02}. 
$H^1$ is the usual Sobolev space of functions that are square integrable on
$\Omega$, whose derivatives are also square integrable.

The discrete systems of equations are derived by formally expanding 
the test and trial functions in terms of a basis $\{\phi_i\}$ for $X^N\!.$
Consider the weak form of the Poisson equation, 
{\em Find $u\in X_0^N$ such that
$(\nabla v,\nabla u)=(v,f)$ for all $v \in X_0^N,$} with
\begin{eqnarray}
\nonumber \\[-2ex]
  u(\bx)=\sum_{j=1}^n u_j \phi_j(\bx).
\end{eqnarray}
Inserting this expansion and taking $v=\phi_i$, we get 
\begin{eqnarray} \label{eq:aip}
    A \uu &=&  \ub,
\end{eqnarray}
with $\uu=[u_1\dots u_n]^T$ the vector of unknown basis coefficients
and
\begin{eqnarray} \label{eq:A}
   A_{ij} := (\nabla \phi_i,\nabla \phi_j),
\end{eqnarray}
the symmetric positive definite (SPD) stiffness matrix associated with the
Poisson operator.  Elements of the data vector $\ub$ are computed by
evaluating inner products $b_i := (\phi_i,f)$.

\begin{figure}
{\setlength{\unitlength}{1.0in}
   \begin{picture}(3.000,0.90)(-.09,.05)
      \put(-.04,0.00){\includegraphics[width=2.9in]{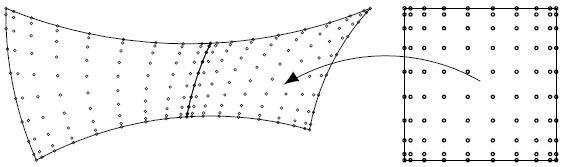}}
      \put(2.37,0.23){\Large $\Oh$}
      \put(1.25,0.30){\large $\Omega^e$}
      \put(0.20,0.19){\large $\Omega^{\hat e}$}
   \end{picture}}
\caption{\label{fig:map2d} \small
$9$th-order SE mapping from canonical domain $\Oh$ to physical 
subdomain $\Omega^e \subset \RR^2.$
\vspace*{-.2in} }
\end{figure}

To derive fast matrix-free operator evaluations for iterative solution 
of (\ref{eq:aip}), we introduce the local SE basis functions.
To begin, we assume 
$\Omega = \cup_{e=1}^E \Omega^e$, where the non-overlapping 
subdomains (elements) $\Omega^e$ are images of the reference 
domain, $\br \in \Oh = [-1,1]^3$, given by 
\begin{eqnarray}
\hspace*{-.2in}
\nonumber \left.  \bx \right|^{}_{\Omega^e} \!\!  &=&  \! \!  \bx^e(r,s,t)\\
\label{eq:xijk} \!\!  &=& \! \!  \sum_{k=0}^N\sum_{j=0}^N\sum_{i=0}^N
\bx^e_{ijk} \, h_i(r) \, h_j(s) \, h_k(t),
\end{eqnarray} 
as illustrated in Fig. \ref{fig:map2d}.
Here, $h_i(r)$ ($s$, or $t$) $\in \PP_N$ is assumed to be a
cardinal Lagrange polynomial, $h_i(\xi_j)=\delta_{ij},$ based on the
Gauss--Lobatto--Legendre (GLL) quadrature points, $\xi_j \in [-1,1]$,
$j=0,\!\dots\!,N\!.$  This choice of points yields well-condi\-tioned 
operators and allows for accurate pointwise quadrature with a 
diagonal mass matrix. 

All functions in $X^N$ have a form similar to (\ref{eq:xijk}).  
For example, the scalar 
$\left. u(\bx)\right|^{}_{\Omega^e} = u(\bx^e(\br)) =: u^e(\br)$
is written in terms of the $(N+1)^3$ local basis coefficients 
$\uu^e := \{ u^e_{ijk}\}$,
\begin{eqnarray} \label{eq:uijk} 
 \left.  u \right|^{}_{\Omega^e} =
   \sum_{k=0}^N\sum_{j=0}^N\sum_{i=0}^N u^e_{ijk}\,h_i(r)\,h_j(s)\,h_k(t).
\end{eqnarray} 
An important consequence of the GLL-based tensor-product Lagrange polynomial
representation is that differentiation with respect to $r$, $s$, and $t$ at
quadrature points $\bxi_{\ih \jh \kh}=( \xi_\ih, \xi_\jh, \xi_\kh)$ can be
expressed as efficient tensor contractions.  Let
\begin{eqnarray} \label{eq:dh} 
\Dh_{\ih i} & := & \left. \dd{h_i}{r} \right|^{}_{\xi_{\ih}}
\end{eqnarray}
be the one-dimensional differentiation matrix mapping from the nodal 
points to the (identical) quadrature points, and let $\Ih$ be the
$(N\!+\!1)\times(N\!+\!1)$ identity matrix.  Let $\uu^e_r$, $\uu^e_s$,
$\uu^e_t$ denote partial derivatives of $u^e(\br)$ with respect to
the coordinates $\br=(r,s,t)=(r_1,r_2,r_3)$ evaluated at the GLL points.
Then
\begin{eqnarray} \label{eq:gradr} 
\hspace*{-.32in}
\uu^e_r = D_1 \uu^e := (\Ih \otimes \Ih \otimes \Dh)\, \uu^e 
              =  \sum_{\ih} \Dh_{i \ih} u^e_{\ih j k},  \;\;
  \\ \label{eq:grads} 
\hspace*{-.32in}
\uu^e_s = D_2 \uu^e := (\Ih \otimes \Dh \otimes \Ih)\, \uu^e 
              =  \sum_{\jh} \Dh_{j \jh} u^e_{i \jh k},  \;\;
  \\ \label{eq:gradt} 
\hspace*{-.23in}
\uu^e_t = D_3 \uu^e := (\Dh \otimes \Ih \otimes \Ih)\, \uu^e 
              =  \sum_{\kh} \Dh_{k \kh} u^e_{i j \kh}.  
\end{eqnarray} 
The chain rule is used to differentiate with respect to
$\bx=(x,y,z)=(x_1,x_2,x_3)$:  
\begin{eqnarray} \label{eq:grad}
\hspace*{-.3in}
\nabla^e \uu^e = \bD^e \uu^e =
\left.  
    \pp{u^e}{x^e_p} 
\right|_{\bxi_{ijk}}^{} \!\!\! = \,
\sum_{q=1}^3
\left(
    \pp{r_q}{x^e_p}
    \pp{u^e}{r_q} 
\right)^{}_{\bxi_{ijk}} \!,
\end{eqnarray}
where derivatives with respect to $r_q$ are computed by
(\ref{eq:gradr})--(\ref{eq:gradt}).  The array of metrics $\pp{r_q}{x^e_p}$ is
found by inverting (at each grid point, $\bxi_{ijk}$) the $3 \times 3$ matrix,
$\pp{x^e_p}{r_q} = D_q \ux^e_p$.  

Note that evaluation of the full gradient
(\ref{eq:grad}) involves three tensor contractions, 
(\ref{eq:gradr})--(\ref{eq:gradt}), each requiring $2(N+1)^4$ operations 
and $(N+1)^3$ memory references, followed by three pointwise contractions,
$u^e_{x_p} = \sum_{q=1}^3 r_{q,x^e_p} \, u^e_{r_q}$, requiring 
15$(N+1)^3$ operations and $9(N+1)^3$ memory references (assuming that
$u^e_{r_q}$ is cached from the $\nabla_r$ operation).  For the full
field, the gradient $(\uu_x,\uu_y,\uu_z) \longleftarrow \uu$ thus requires
$10 E(N+1)^3 \approx 10n$ memory references and $
6E(N+1)^4 + 15E(N+1)^3 \approx n(15+6N)$ operations.  
We note that the $O(N^4)$ work terms (\ref{eq:gradr})--(\ref{eq:gradt})
are readily cast as
dense matrix-matrix products \citep{tufo99a,dfm02}.

In addition to differentiation, the weighted residual formulation
requires integration, which is effected through GLL quadrature.
For any $u,v \in X^N$, we define the discrete inner products,
\begin{eqnarray} \label{eq:qN}
\hspace*{-.3in}
\left( v,u\right)^{}_N\!&:=&\!\sum_{e=1}^E \left(v,u\right)^e , \\ \label{eq:qe}
\hspace*{-.3in}
\left( v,u\right)^{e}\!&:=&\!
\sum_{k=0}^N
\sum_{j=0}^N
\sum_{i=0}^N 
v^e_{ijk} \rho_{ijk} \cJ^e_{ijk} u^e_{ijk} \\ \label{eq:vBu}
\!&
=
&\!
(\uv^e)^T B^e \uu^e.
\end{eqnarray}
Here, 
$\rho_{ijk} = \rho_i \rho_j \rho_k$ is the product of one-dimensional
 GLL quadrature weights;
 $\cJ^e(\br) = \left| \pp{x^e_p}{r_q} \right|$ is 
the Jacobian associated with the map $\bx^e$ from $\Oh$ to $\Omega^e$;
and $B^e$=diag$(\rho_{ijk} \cJ^e_{ijk})$ is a local, diagonal mass matrix.
For affine maps, $(v,u)^e \equiv \int_{\Omega^e} v u \, dV$ whenever
the product $vu$ is a polynomial of degree $2N-1$ or less.  
The high accuracy realized by the GLL quadrature is sufficient to ensure
stability for the Poisson operator when $(\nabla \phi_i,\nabla \phi_j)_N$
replaces $(\nabla \phi_i,\nabla \phi_j)$ in (\ref{eq:A}) \citep{ronpa871}, but
not for the advection operator, where the integrand is of degree 3$N$
and thus requires higher-order integration \citep{johan13}.

Equipped with the basic calculus tools (\ref{eq:gradr})--(\ref{eq:vBu}), we
evaluate the bilinear form $(\nabla v,\nabla u)$ as follows,
\begin{eqnarray} \label{eq:form}
\hspace*{-.3in}
(\nabla v,\nabla u) &=& \sum_{e=1}^E (\nabla v^e,\nabla u^e)^e \\ \nonumber
  &=& \sum_{e=1}^E (\uv^e)^T A^e \uu^e \;=\;
  \uv_L^T A_L \uu_L,
\end{eqnarray}
where 
$\uu_L=[\uu^1\cdots \uu^E]$ is the collection of all local basis vectors
and $A_L$=block-diag($A^e$) comprises the local stiffness matrices given 
in factored form by
\begin{eqnarray} \label{eq:Ae}
\hspace*{-.3in}
A^e =
\left(\!  \! \begin{array}{c}
D_1 \\[.3ex]
D_2 \\[.3ex]
D_3 \end{array} \!  \right)^{    \! T}   \!\! \! \!
\left(\!  \begin{array}{ccc}
G^e_{11} & G^e_{12} & G^e_{13} \\[.3ex]
G^e_{12} & G^e_{22} & G^e_{23} \\[.3ex]
G^e_{13} & G^e_{23} & G^e_{33} \end{array} \! \right)    \! \!
\left( \! \! \begin{array}{c}
D_1 \\[.3ex]
D_2 \\[.3ex]
D_3 \end{array}\! \!  \right).
\end{eqnarray}
Here, the six local geometric factors $G^e_{ij}=G^e_{ji}$ are diagonal
matrices with one nontrivial entry for each gridpoint $\bxi_{ijk}$,
\begin{eqnarray} \label{eq:gij}
\left[ G^e_{mm'}  \right]^{}_{ijk} =
\left[ \sum_{l=1}^3 \pp{r_m}{x^e_l}\pp{r_{m'}}{x^e_l} \right]^{}_{ijk}
\!\!\!
B^e_{ijk}.  
\end{eqnarray}
Evaluation of $A^e \uu^e$ thus requires $7(N+1)^3$ memory references
(six for $G^e_{mm'}$ and one for $\uu^e$) and $12(N+1)^4 + 15(N+1)^3$
operations, per element.
Aside from preconditioning, $A^e \uu^e$, {\em constitutes 
the principal work for the pressure-Poisson problem and for the viscous
solves.}  In the case of the Jacobi-preconditioned viscous substeps,
it is the only work term that scales as $O(nN)$.   The other contributions
in Jacobi PCG total to $\sim$ 12$n$ floating-point operations.

To ensure interelement continuity ($u,v \in X^N \subset H^1$), one must
constrain local basis coefficients at shared element interfaces to be equal.
That is, for any given sets of coefficient indices $(i,j,k,e)$ and
$(\ih,\jh,\kh,\eh)$,
\begin{eqnarray} \label{eq:dssum} 
   \bx_{ijk}^e = \bx_{\ih \jh \kh}^{\eh}  &
               \longrightarrow & u_{ijk}^e = u_{\ih \jh \kh}^{\eh}.  
\end{eqnarray} 
The statement (\ref{eq:dssum}) leads to the standard finite element procedures
of matrix assembly and assembly of the load and residual vectors.  
Matrix-free algorithms that use iterative solvers require assembly only of
vectors, since the matrices are never formed.  

To implement (\ref{eq:dssum}), we introduce a {\em global-to-local map},
formally expressed as a sparse matrix-vector product, $\uu_L\! =\! Q \uu$,
which takes global (uniquely defined) degrees of freedom $u_l$ from the index
set $l \in \{1,\!\dots\!,n\}$ to their (potentially multiply defined) local
counterparts $u_{ijk}^e$.  The continuity requirement leads to
\begin{eqnarray} 
\hspace*{-.0in}
\uv_L^T \!A_L \uu_L =
\uv^T Q^T\! A_L Q \uu =
\uv^T\! A \uu,
\end{eqnarray} 
from which we conclude that the global stiffness matrix is $A=Q^T \!A_L Q.$
We refer to
$A_L$ as the unassembled stiffness matrix and $A$ as the assembled
stiffness matrix.\footnote{We typically denote
$Q^TA_L Q=:\Ab$ as the {\em Neumann operator}, which is orthogonal
to the constant vector, and $A=R\Ab R^T$ as the 
SPD stiffness matrix, where $R$ is a restriction matrix that 
discards rows corresponding to Dirichlet data \cite{dfm02}.
Application of $R$ does not impact complexity so we do not
discuss it further.}
With this factored form, a matrix-vector product can be
evaluated as $\uw = Q^T\! A_L Q \up$, which allows parallel evaluation of the
work-intensive step of applying $A^e$ to basis coefficients in each element
$\Omega^e$.  Application of the Boolean matrices $Q$ and $Q^T$ represents the
communication-intensive phases of the process.\footnote{In the case of
nonconforming elements, $Q$ is not Boolean but can be factored into a Boolean
matrix times a local interpolation matrix \citep{dfm02}.}
We note that $Q$ and $Q^T$ are the spectral element/finite element emthod equivalents of the finite difference
``halo'' exchange.   Unlike finite differences, however, $Q$ and $Q^T$ 
have a {\em unit-depth stencil for all $N$, and the discretization is thus
communication minimal}.

\section{Parallel GPU Development}

Our parallel approach to solving the incompressible NS equations follows the
standard SPMD paradigm of partitioning the domain across $P$ MPI ranks, each
with its own private address space, and time advancing the equations in a
cooperative fashion using iterative solvers to solve the elliptic subproblems
for the velocity, temperature, and pressure.   On each node, we run one MPI
rank per GPU.  All data resides on the device, with a copy back to the host
only when needed (e.g., for I/O or analysis of turbulence statistics).

NekRS has been developed in close collaboration with the libParanumal project,
\cite{warburton2019,warburton2019b,ChalmersKarakusAustinSwirydowiczWarburton2020,streamParanumal2020}
which provides high-performance kernels for high-order methods on GPUs.  The
GPU kernels are written in the portable Open Concurrent Compute Abstraction
(OCCA) library \cite{medina2015okl,occa,occa-web} to abstract between different
parallel 
languages such as OpenCL, CUDA, and HIP. OCCA allows developers to implement 
the parallel kernel
code in a slightly decorated C++ language, OKL.  At runtime, the user can specify
which parallel programming model to target, after which OCCA translates the OKL
source code into the desired target language and Just-In-Time (JIT) compiles
kernels for the user's target hardware architecture.  In the OKL language,
parallel loops and variables in special memory spaces are described with simple
attributes. For example, iterations of nested parallel for loops in the kernel
are annotated with \texttt{@outer} and \texttt{@inner} to describe how they are
to be mapped to a grid of work-item and work-groups in OpenCL or threads and
thread-blocks in CUDA and HIP. All iterations that are annotated with
\texttt{@outer} or \texttt{@inner} are assumed to be free of loop carried
dependencies.  We describe several of the kernels in detail below.

We note that experience with Mira \citep{fischer15} has
established that the strong-scale limit for Nek5000 is around $n/P=2000$--4000
points per MPI rank (in -c32 mode), meaning using just 2 to 8 elements per rank
for typical orders of $N$=7 to 11.  By contrast, the strong-scale limit on
modern GPUs such as the Nvidia V100 is around $n/P=2$--4 million points per
rank (i.e., per GPU), or about 4,000 to 8,000 elements \citep{fischer20a}.
Moreover, ``hero'' runs on Mira were at the level of about 15 million elements,
which could easily be run on a million ranks.  On Summit, 
NekRS routinely is run with 175 million elements and $N=7$ ($n=60$B).  The
quantitative differences have an impact on considerations such as communication
hiding and the importance of internode latency. We discuss these issues further
in Section~\ref{perform}.

\subsection{Domain Partitioning}

For the large runs that are now routine on Summit, we partition the domain
using parallel recursive spectral bisection (parRSB) \citep{pothen90} in such a
way that the number of elements on each processor differs by at most 1.   
The Fiedler vector for parRSB is computed by using
either restarted Lanczos or inverse iteration with lean algebraic multigrid
(AMG) \citep{livne2012} on the element-centered connectivity graph.   Prior to
running parRSB, we execute recursive coordinate bisection (parRCB) in order to
organize the graph into reasonably connected subsets on each processor.
Otherwise, the parRSB iterations can incur significant communication overhead
because one or more processors may have each of its elements connected to
different processors if the ordering is arbitrary (e.g., partitioned according
to the original element numbering).  Prepartitioning with parRCB can cut
parRSB run times by a factor of 100.  On GPU-based systems parRCB/RSB
are run on the CPUs because the number of elements,
$E$, in the SEM is typically three orders of magnitude smaller than the number of
grid points, $n=EN^3$.  On 8,100 cores of Summit, the partition time for $E=60$
million is about 40 seconds using Lanczos with a parRCB preprocessing time of
0.8 seconds.

\subsection{Parallel Communication}

Time advancement of the discretized NS equations effectively amounts
to executing a sequence of matrix-vector products.  Because both the trial and
test spaces ($\bu,p$ and $\bv,q$) are continuous, these products
involve a map of the form $\uw = Q^T\! Z_L Q \uu$, where
$Z_L$=block-diag($Z^e$) represents some localized physics (e.g., advection or
diffusion), $Q^T$ reflects the continuity of the test functions, and $Q$ the
continuity of the trial functions.  When recast as $\uw_L = QQ^T\! Z_L \uu_L$,
the parallelization is clear:  each processor evaluates ${\tilde \uw}^{e} = Z^e
\uu^e$, $e=1,\dots,E_p$, where $E_p$ is the number of elements on rank $p$ 
(With the private memory model, other discriminators are not required for $Z^e$
or $\uu^e$ because their data is implicitly indexed by $p$.)
This parallel work step is followed by the communication phase, $\uw_L = QQ^T
{\tilde \uw}_L$, which corresponds to an exchange and sum of shared interface 
values between adjacent elements.  On Mira, $QQ^T$ is almost 100\% communication
(latency) dominated in the strong-scale limit of one or two elements per rank.
On GPUs there  are thousands of elements per rank. A significant portion of
the elements are thus interior to the local partition, and there is consequently
an opportunity to overlap work and communication.

$QQ^T$ is implemented in parallel by using the open-source communication
library {\em gslib}, which supports multiple data types and
associative/commutative 
operations (e.g., $\min$, $\max$, $*$, $+$) for scalar and vector fields, 
as well as one-sided operations $Q$ and $Q^T$.  
   The adjacency graph is prescribed by a simple interface:  
The user provides indices in a vector, $\ug_{p}$ of length
$n_p = E_p(N+1)^3$ on each processor, $p=0,\dots,P-1$.
The indices correspond to global pointers, while
their positions, $1,\dots,n_p$ correspond to local pointers.  
Passing $\ug_p$ to {\em gs\_setup}
returns a handle, {\em gsh}, which is then used when executing
{\em gs\_op(gsh,$\uw_L$,+)} to produce $\uw_L \longleftarrow QQ^T\uw_L$.
The user does not need to know which processor holds the adjacent elements
or anything else about the shared indices.  If a global pointer is 
unique in the set $\bigcup_p \{ \ug_p \}$, then the corresponding
entry in $\uw_L$ will be unchanged.  If one knows a priori that
certain entries in $\ug_p$ are singletons, a 0 index may be supplied
in $\ug_p$, which saves work in the discovery phase of {\em gs\_setup}.
(We typically set all element-interior pointers 0.)

   At setup time, {\em gslib} picks a communication strategy (pairwise,
crystal router\footnote{Crystal router is a scalable generalized all-to-all
\citep{fox88}.} or all-reduce) that yields the lowest maximum time over a set
of trials for the given adjacency graph.  We have found this optimization
to be particularly important (10$\times$) in the context of 
AMG for solution of distributed coarse-grid problems \citep{fischer08a}.  
When the number of nonzeros per row is large, the pairwise exchange can require
a large number of messages, whereas crystal router requires only $\log_2 P$
messages.  The all-reduce approach has a nominal $\log_2 P$ cost; but with
hardware support, the cost is effectively independent of $P$ and (on Mira)
bounded by 4$\times$ the latency for short messages, even with $P > $1M.
{\em gs\_setup}  executes in $O(\log P)$ time and is quite fast.  For example,
with 3B gridpoints on a million ranks of Mira, the time (including setup
and 10 trials each for pairwise and crystal router) is less than one second.
For this example, the corresponding {\em gs\_op} times are .000325 seconds for
pairwise and .0046 seconds for the crystal router.

For device-based implementations of $QQ^T$, we extend the autotuning approach
to test device-to-device transfers (i.e., GPU direct), transfers via the host
with buffers packed on the host, and transfers via the host with buffers packed
on the GPU.  Since overlapped communication and computation is also supported,
the determination of the fastest algorithm requires testing under overlapped
run conditions, which is enabled through a call-back function that allows the
setup to be tested in tandem with execution of the relevant kernel.   The
overlap works as follows. All elements with nonlocal adjacency connections
are evaluated first (i.e., ${\tilde \uw}^{e_{nl}} = Z^{e_{nl}} \uu^{e_{nl}}$, where
$e_{nl}$ spans the set of elements having nonlocal connections).  The nonlocal
communication is initiated, and the remaining local products are evaluated.
Incoming off-device contributions are then added to the result, $\uw_L$. 

\begin{figure}
 \includegraphics[width=0.45\textwidth]{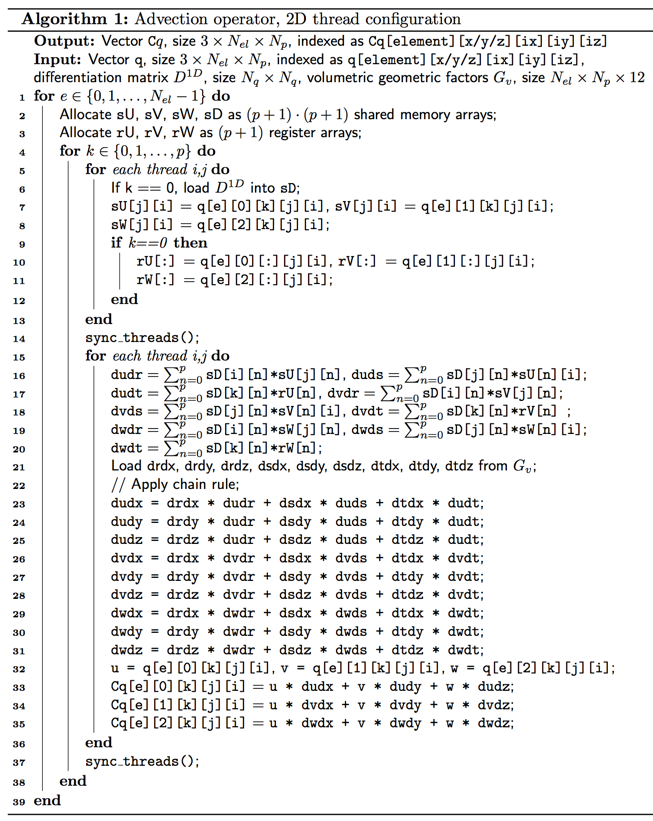}
 \caption{\label{fig:alg1} Two-dimensional thread-block for advection.
\vspace*{-.2in}
}
\end{figure}

\subsection{High-Order Kernels}

\begin{figure}
 \includegraphics[width=0.45\textwidth]{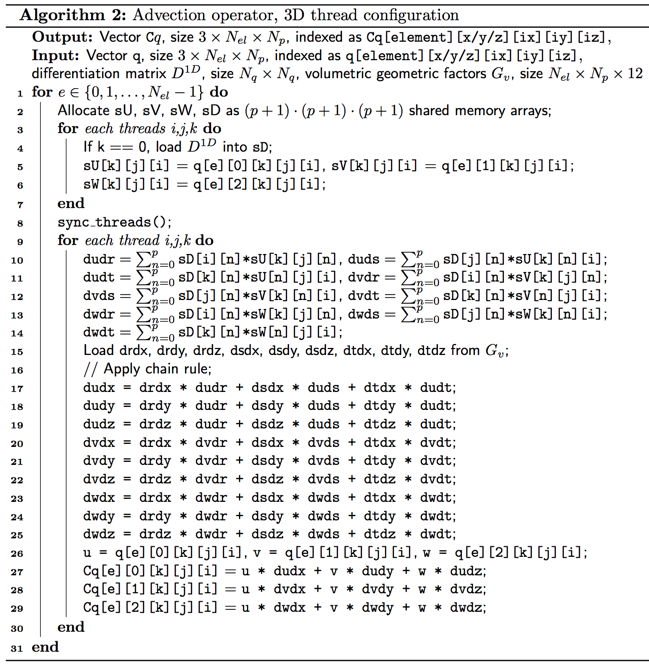}
 \caption{\label{fig:alg2} Three-dimensional thread-block for advection.
\vspace*{-.2in}
}
\end{figure}

The $O(N)$ computational intensity of the spectral element method,
coupled with minimal indirect addressing, provides significant performance
opportunities on GPU architectures.  Moreover, for vector- (e.g., velocity-)
oriented operations, the $O(N^3)$ geometric factors associated with each
element can be reused across each velocity component (e.g., when computing
$\nabla \bu$ on $\Omega^e$).  
  While the majority of NekRS simulations are run with $N$=7, it is not uncommon
to require $N$=11--15 for some applications, implying that the dealiased
advection operators will be evaluated on $N_q^3$ quadrature points, with 
$N_q=$ 13--23.  For these cases, a 2D thread structure must be used, 
as illustrated in Fig. \ref{fig:alg1}.
  For relatively low $N$, operations may be organized into 3D thread structures
and still be within shared-memory and thread-block limits, 
as illustrated in Fig. \ref{fig:alg2}.

\newcommand{\var}{\texttt}

NVIDIA imposes a
hard limit of 1,024 threads per thread block, meaning that a triply nested 3D
thread structure of size $N_q \times N_q \times N_q$ mandates $N_q \leq 10$.
Therefore, the maximum achievable polynomial order in the 3D thread
structure advection operator, illustrated in Algorithm 2, is $N=9$.  
The 2D thread structure of Algorithm 1, however, does not reach
the 1,024 threads per threadblock restriction until $N_q = 32$.  Further, using
a 2D thread structure allows for the shared-memory usage to be better
optimized.  For example, in Algorithm 1, only four 2D
shared-memory structures are needed.  This approach is already available
in libParanumal \cite{warburton2019}.  A small development building on this
earlier work is the addition of using 2D shared-memory structures with a 1D
register memory structure.  Since each thread $(i,j)$ has its own register
data, array lookups of the form $\var{A[*][j][i]}$ can be reduced into 
one-dimensional register array lookups of the form $\var{rA[*]}$, where $\var rA$
is a register array for thread $(i,j)$ and $\var{rA[k]=A[k][j][i]}$
corresponds to values of $\var A$ along the $k$-index for a fixed $(i,j)$.  A
major advantage of this approach is reducing the number of (relatively) slow
shared-memory loads by nearly a factor of three.  In addition, this helps
preserve the 48 kB of shared memory available on the NVIDIA V100, which
can hold only 6,144 double-precision words.  Because OCCA allows for runtime
JIT-compilation of kernels,` either the 2D or 3D thread structure kernel may be
used for the case $N_q\le 10$, based on which is more performant.

\subsection{Poisson Solve Preconditioning}

As noted above, the pressure-Poisson solve is intrinsically the stiffest
substep in NS time advancement.  It is easy to understand why this is
so by considering the example of flow in a pipe.  
When flow is suddenly forced
in at a given flow-rate on one end of the pipe, it must leave at the same
rate at the far end of the pipe and must have the same mean value throughout
the pipe---all by the end of the current timestep.  
A consequence of the divergence-free constraint is that the
Poisson problem is intrinsically communication intensive---all processors
must ``know'' about an inlet condition, which might be prescribed only
on one processor.
Fast scalable Poisson {\em solvers} are thus of paramount concern for
incompressible simulations.  Because the setup is amortized over tens of
thousands or hundreds of thousands of timesteps, we care more about {\em solve}
times than {\em setup} costs.

For CPU-based applications, we have developed a $p$-multigrid strategy 
that uses an overlapping additive Schwarz method (ASM) as a smoother
\cite{fischer04,lottes05}.  
The local solves are effected in $\approx 12E(N+3)^4$ operations by using
tensor-contraction-based fast-diagonalization methods (FDMs) \cite{dfm02}.
Typical multigrid schedules use approximation
orders $N$, $N/2$, and $N=1$ at successively coarser levels.  
For modest-sized meshes (e.g., $E < 500$K), the $O(E)$-sized coarse-grid
problem is solved by using a fast direct solver that requires a minimal $2 \log_2
P$ message exchanges \cite{tufofisc01}.  For larger problems on CPU platforms,
the coarse-grid problem is solved by using communication-minimal implementations
of algebraic multigrid.  Here, the adaptability of {\em gslib} is
essential because of the stencil growth in the lower levels of AMG.  At each
level, communication is effected by the fastest supported algorithm in {\em
gslib}, resulting in a 5- to 10$\times$ improvement over simply using pairwise
exchanges for stencil updates \cite{fischer08a}.
Our CPU implementation of ASM also uses an additive approach {\em between}
levels, which means that there is only one proper matvec in $A$ per PCG
iteration.  The idea is that each element of the Krylov subspace should be
projected, rather than using cycles for unprojected iterations.  On BG/Q, which
has hardware support for all-reduce ($< 20 \mu s$ for $P=1$M \cite{fischer15}),
dot products for projection incur only a fraction of a percent of total run
time.
We use flexible PCG because weighting the ASM, which improves its
smoothing properties, introduces a slight asymmetry in the preconditioner.

\begin{figure}
{\setlength{\unitlength}{1.0in} \begin{picture}(3.000,1.50)(.0,.0)
  \put(0.00,0.00){\includegraphics[width=3.00in]{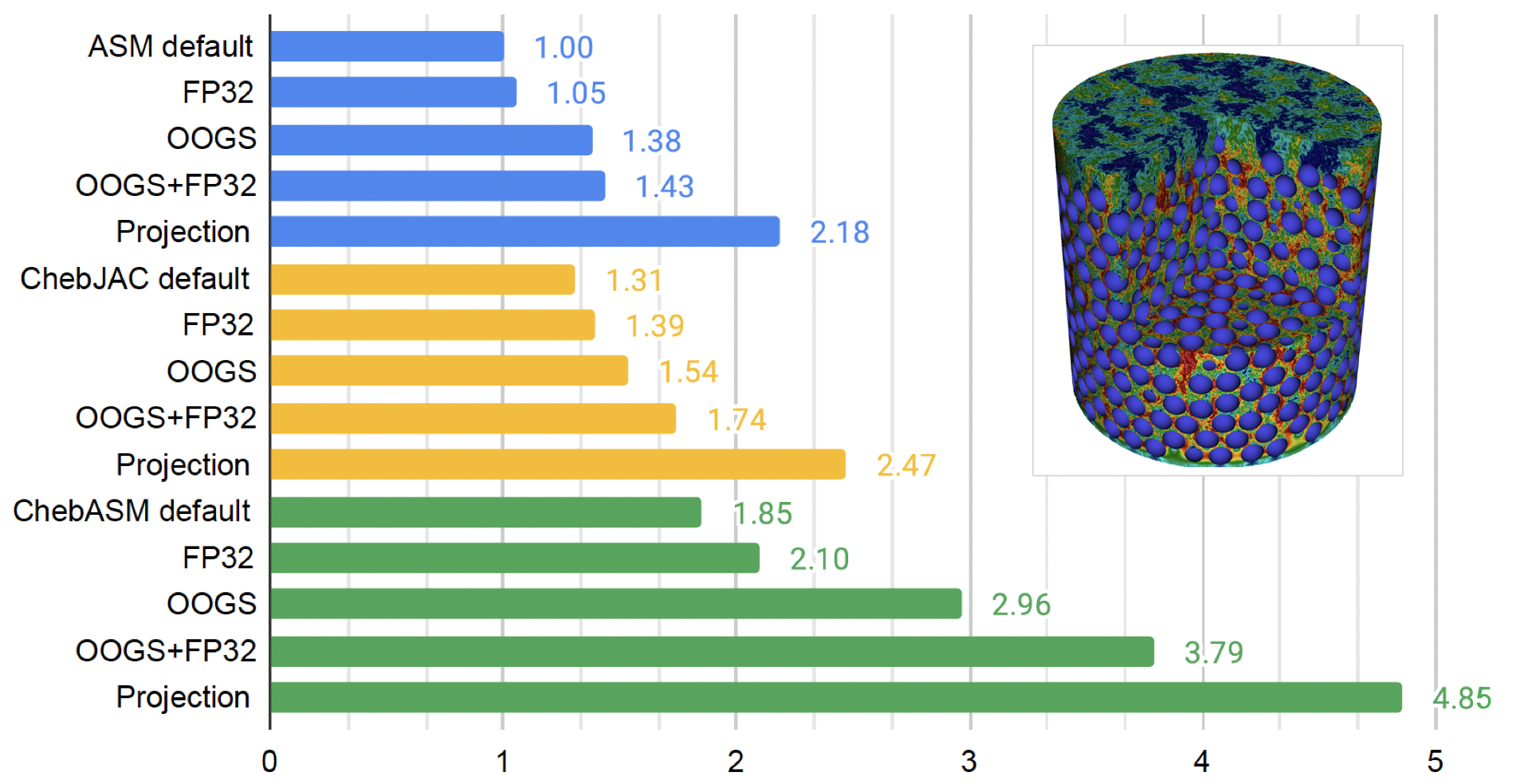}}
\end{picture}}
\caption{\label{fig:pb1568_base} \small
Successive gains in pressure-solve performance for ASM, CHEBY-JAC, and
CHEBY-ASM $p$-multigrid smoothers as a function of implementation options for
the 1568-pebble case ($E=524$K, $N=7$, characteristics) on 22 nodes of Summit
($n/P=1.36$M, $P=132$ V100s).  \vspace*{-.2in} }
\end{figure}

\begin{table*} [t]
  \footnotesize
  \begin{center} \begin{tabular}{|c|l|c|c|c|c|c|c|c|c|c|}
  \hline
  \multicolumn{11}{|c|}{{\bf NekRS Preconditioning Development, $n=2,569,495,663$}}\\
  \hline
   Timestepper & Smoother    & GPU &  $E$ & $N$ & $n$/GPU & $\Delta t$ & CFL & $v_i$ & $p_i$ & $t_{step}$ (s) \\
  \hline
              &    RAS       &  54 & 524386 & 7 &  3.33M &  1.0e-03 & 4.05 &  4  & 112 & 1.64e+00   \\
              &    ASM       &  54 & 524386 & 7 &  3.33M &  1.0e-03 & 4.05 &  4  &  93 & 1.44e+00   \\
   CHAR-BDF2  & CHEBY-JAC    &  54 & 524386 & 7 &  3.33M &  1.0e-03 & 4.05 &  4  &  32 & 1.04e+00   \\
              & CHEBY-RAS    &  54 & 524386 & 7 &  3.33M &  1.0e-03 & 4.05 &  4  &  26 & 6.56e-01   \\
              & CHEBY-ASM    &  54 & 524386 & 7 &  3.33M &  1.0e-03 & 4.05 &  4  &  16 & 5.03e-01   \\
  \hline
              &    RAS       &  54 & 524386 & 7 &  3.33M &  2.5e-04 & 1.06 &  2  &  51 & 7.54e-01   \\
              &    ASM       &  54 & 524386 & 7 &  3.33M &  2.5e-04 & 1.06 &  2  &  39 & 6.38e-01   \\
   BDF3-EXT3  & CHEBY-JAC    &  54 & 524386 & 7 &  3.33M &  2.5e-04 & 1.06 &  2  &  15 & 4.96e-01   \\
              & CHEBY-RAS    &  54 & 524386 & 7 &  3.33M &  2.5e-04 & 1.06 &  2  &  13 & 3.43e-01   \\
              & CHEBY-ASM    &  54 & 524386 & 7 &  3.33M &  2.5e-04 & 1.06 &  2  &   8 & 2.59e-01   \\
  \hline
  \end{tabular}
\end{center}
\caption{\label{peb1568-precon}
NekRS preconditioner performance comparison on 54 GPUs of Summit for the case
of Fig.~\ref{fig:pb1568_base}.  
A restart file at convective time $t$=20 is used to
provide a turbulent initial condition.  Here, the Courant number (CFL), time
per step in seconds ($t_{step}$), velocity iteration count ($v_i$), and
pressure iteration count ($p_i$) are all averaged over 100 steps.}
\end{table*}

On GPUs, the situation is somewhat different.  First, the arithmetic for the
matvecs and local solves is very fast.  Because of other strong-scaling
limitations, there are enough elements per rank (typically, $E/P \sim
4000$--$8000$ on the NVIDIA V100 and AMD Instinct\textsuperscript{TM} MI100) to effectively overlap
communication with computation on the fine-grid matvecs. Moreover, dot products
are not fast on GPUs.   A better smoother than straight ASM is consequently
more effective.  We consider two strategies.  The first smoother uses
two Chebyshev-accelerated Jacobi (CHEBY-JAC) sweeps, with pre- and
postsmoothing at the top two levels.  Hypre or parAlmond
\cite{gandham2014,remacle2016} is used for the
$O(E)$ coarse-grid problem.  (AMG cannot be applied directly to the dense
$A\uu=\ub$ systems.  It can, however, be applied to FEM-based surrogates for
$A$ to develop an alternative preconditioning strategy
\cite{sao80,canuto10,pedro19}.)
The second (CHEBY-ASM) applies Chebyshev acceleration to the
FDM-based ASM smoother, with the coarse-grid solver unchanged.

The effectiveness of CHEBY-JAC is illustrated in Fig.~\ref{fig:pb1568_base}, 
which contrasts with the baseline ASM results under a succession of
algorithmic refinements, including switching to FP32\footnote{FP32 
is used only for the smoothing steps.  Indirect-addressing
overheads result in minimal gains if FP32 is used for the unstructured
coarse-grid solve.} for certain parts of the
preconditioner and/or using an adaptive gather-scatter that chooses between
different (device/host) buffer packing and pairwise exchange strategies,
and incorporating projection-based initial guesses \cite{fisc98}.  
The test problem is turbulent flow through a cylinder with 1,568 spherical
pebbles at $Re_D=5000$, run on 22 nodes of Summit.  The discretization consists
of $E=524$K elements of order $N=7$ ($n=180M$), with $n/P=1.36$M, (i.e.,
beyond the 80\% strong-scale limit).  Timestepping is based on two-stage
2nd-order characteristics with CFL=4.

\begin{figure}
 \includegraphics[width=0.45\textwidth]{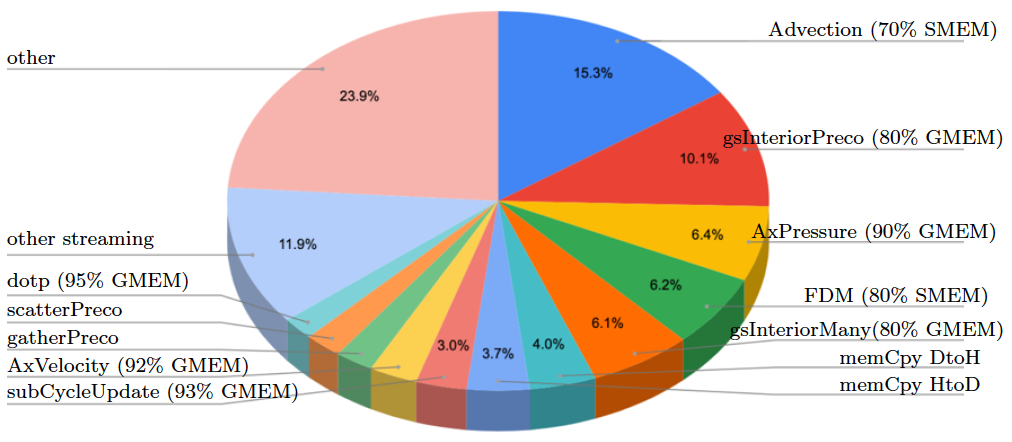}
\caption{\label{fig:pb1568_pie} \small
 Timing breakdown for the case of Fig.~\ref{fig:pb1568_base}.
 }
\end{figure}

Figure~\ref{fig:pb1568_base} shows that the
straight CHEBY-JAC strategy (yellow) is 30\% faster than baseline ASM (blue). 
With the enhancements, CHEBY-JAC yields a 2.5$\times$ gain over the
baseline.
The CHEBY-ASM case (green) shows similar
improvements (1.85$\times$) in the default configuration but yields an overall
4.85$\times$ speedup in the pressure solve when using the adaptive
gather-scatter, FP32-based smoothing, and projection-based initial guesses.

In addition to ASM, we explored the potential of restricted additive Schwarz
(RAS) \cite{ras99}, in which each subdomain (spectral element) retains its own data
after the FDM solve, rather than exchanging and adding (with counting weight
$\uw$ \cite{lottes05}).  These results, along with the others, are presented for
the 1,568-pebble case on 9 nodes of Summit in Table \ref{peb1568-precon}.
With $n/P > 3.3$M, this case is well above 80\% parallel efficiency, but
the overall relative performance is similar to the 22-node case of
Fig.~\ref{fig:pb1568_base}. 
This table also shows the advantage of the characteristics timestepping,
which allows a $4\times$ gain in stepsize with only a $2\times$ increase
in time per step.

Fig.~\ref{fig:pb1568_pie} shows a performance breakdown of the key
kernels for the CHEBY-ASM results of Fig.~\ref{fig:pb1568_base}. 
Even with the 4.85-fold reduction in the pressure solve time,
about 25\% of the wall-clock time is still directly attributable to
the pressure step, as indicated by the Preco, FDM, and AxPressure
kernels. The interior-element gather-scatter kernels (10\% of wall time)
are limited by and sustain 80\% of global memory bandwidth (GMEM).  The
AxPressure kernel (6.4\%) sustains 90\% GMEM, and the FDM is limited
by and sustains 80\% of shared-memory bandwidth (SMEM).  
For this particular case, the characteristics (\ref{eq:hyp})
accounts for about 18\% of wall time and sustains 70\% of SMEM.



\section{Application Performance}
\label{perform}


In this section, we explore scalability and performance comparisons 
of NekRS for several applications on a variety of platforms.

\subsection{Comparison of Summit and Mira}

\begin{figure} 
 \centering
 \includegraphics[width=0.45\textwidth]{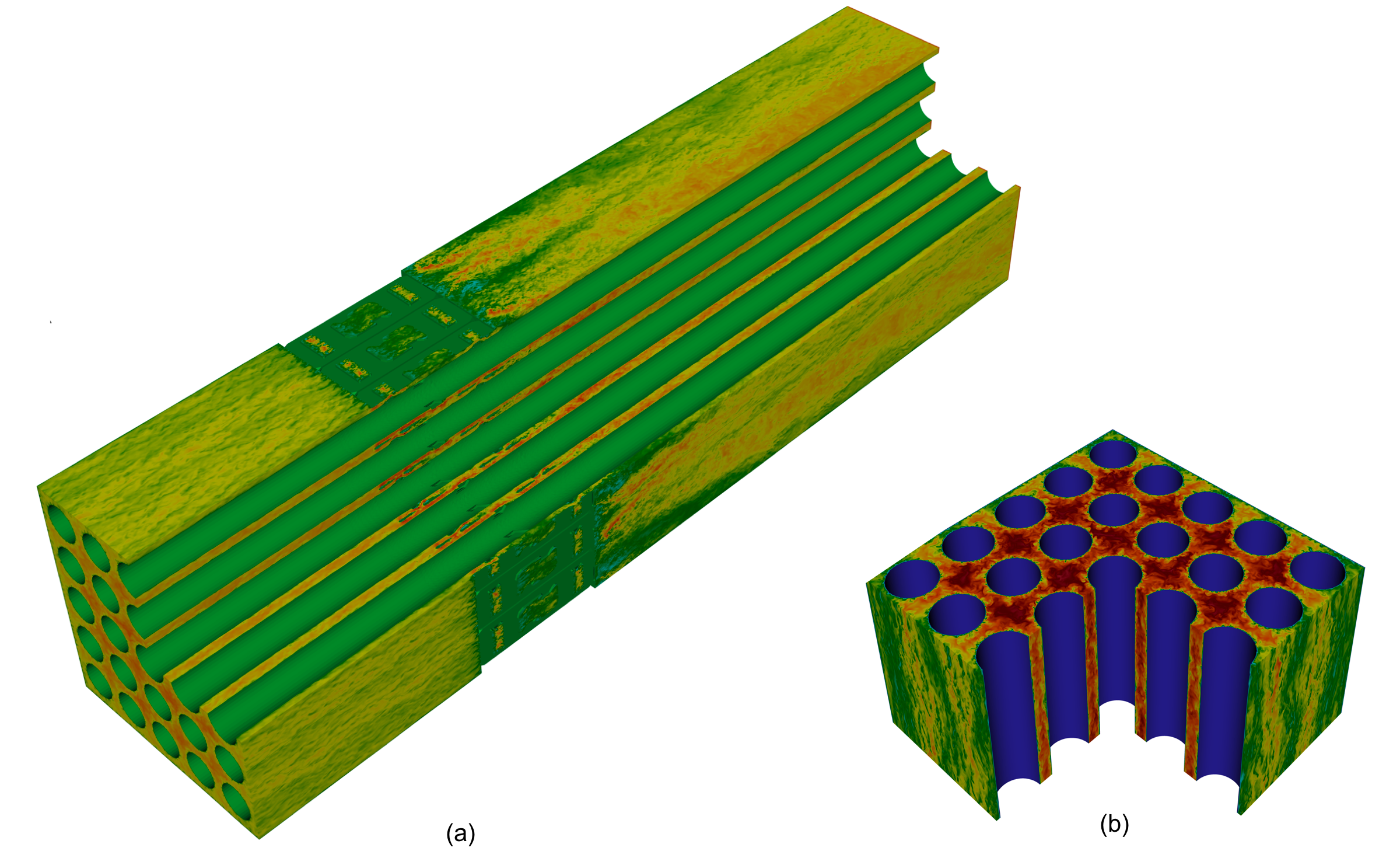}            
 \caption{\label{fig:mira_apps}
  Turbulent velocity snapshots: (a) spacer-grid and (b) DNS 5$\times$5.}
\end{figure}

We begin with comparisons of NekRS on Summit and Nek5000 on Mira for the two
configurations depicted in Fig.~\ref{fig:mira_apps}.  
The first is a 3.2M element LES of a spacer-grid configuration at $Re_D=14,000$,
as considered in \cite{giacomo2019}.  
The second is an 8.4M element DNS in a 5$\times$5 rod bundle configuration at
$Re_D=19,000$ \cite{kraus2020direct}.
Both were run with approximation order $N=7$ on 8,192 nodes of the IBM BG/Q,
Mira, in -c32 mode ($P=262144$ MPI ranks).   For the NekRS runs, we used
tolerances, timestep sizes, and other run parameters that were equivalent to
the Mira-based Nek5000 simulations.   All cases use dealiasing with $N_q=12$.
The only significant algorithmic difference is in the use of CHEBY-ASM for
NekRS versus the default ASM pressure smoother used in Nek5000.   The NekRS
runs used restart files from the Nek5000 cases, and timings were compared over
the same simulation steps.  Table \ref{mira-table} summarizes the comparative
data.

For the spacer-grid case, the number of points per rank on Mira $n/P=4116$,
which corresponds to a parallel efficiency of $\approx$ 80\% (cf. Fig.~1 in
\cite{fischer15}).  The corresponding time per step of $t_{step}=0.68$s
is thus at a minimum for this efficiency.  The average number of ASM-based
pressure iterations per step is 8.7, which constitutes about 27\% of the 
wall-clock time.  This case used characteristics-based timestepping with a
single RK4 substep.  On 98 nodes of Summit we have $t_{step}=0.14$s---4.8
$\times$ faster than Mira at comparable parallel efficiency.   The breakdown of
the Summit GPU wall-clock time in this case is roughly 22\%  for the advection
term (\ref{eq:hyp}), 24\%  for the velocity (viscous) solve, and 52\%  for the
pressure solve.  (The coarse-grid solve time was 8\% of $t_{step}$.)

For the DNS case, the number of points per rank on Mira $n/P=10973$, and
we can assume that 80\% strong-scale limit timings would be realized at
$P \approx 5,242,88$ ranks with $t_{step} \approx 0.35s$.  
In this case, the Summit scaling is not ideal---the best times are at
$n/P=3.9$M, which is significantly larger than what we typically see on Summit.
Whether this anomolous behavior is attributable to system noise or to something
peculiar about the partitioning is as yet unclear.   Nonetheless,
$t_{step}=.183s$ is about $2 \times$ faster than the strong-scale limit times
on Mira.

\begin{table*}[!t]
\newcolumntype{?}[1]{!{\vrule width #1}}
  \footnotesize

  \begin{center} \begin{tabular}{|c|c|c|c|c|c|c|c|c|c|c|c|c|}
  \hline
  \multicolumn{13}{|c|}{{\bf Spacer-Grid, Performance on Mira vs. Summit, $E=3235953$, $N=7$, $n=1.10B$}} \\ 
  \hline
  \hline
  System & Code      & Device & Node & Rank & R/N & $E$/R & $n$/Rank &  $t_{step} (s) $ &  R  & R$_{i}$ & eff \% & R$^{**}$ \\
  \hline
  \hline
   Mira  &Nek5000&CPU &  8192 &262144 & 32 &  12   &  4116   &  6.90e-01 & 1.00 & 1.00 & 100 & 1.00 \\
  \hline          
  \hline          
         &       &    &   38  & 1596  &  42 & 2027  & 695446 &  1.68e+01 &  1.00      &  1  & 100  &0.06  \\
  Summit &Nek5000&CPU &   76  & 3192  &  42 & 1013  & 347723 &  0.50e+01 &  2.12      &  2  & 106  &0.13  \\
         &       &    &  152  & 6384  &  42 & 506   & 173861 &  0.23e+01 &  4.56      &  4  & 114  &0.29  \\
         &       &    &  304  &12768  &  42 & 253   &  86930 &  0.11e+01 &  9.64      &  8  & 120  &0.62  \\
  \hline          
  \hline          
         &       &    &   38  & 1596  &   42 & 2027  & 695446 &   0.78e+01 &  1.00  &  1    & 100  & 0.08 \\
  Summit &NekRS  &CPU &   76  & 3192  &   42 & 1013  & 347723 &   0.38e+01 &  2.01  &  2    & 100  & 0.17 \\
         &       &    &  152  & 6384  &   42 & 506   & 173861 &   0.20e+01 &  3.81  &  4    &  95  & 0.33 \\
         &       &    &  304  &12768  &   42 & 253   &  86930 &   0.11e+01 &  6.72  &  8    &  84  & 0.59 \\
  \hline                                                             
  \hline          
         &       &    &   38  & 228   &  6 &  14193&  4.8M &  2.75e-01 & 1.00 &1.00 & 100 &  2.46 \\
  Summit & NekRS &GPU &   60  & 360   &  6 &  8988 &  3.0M &  1.92e-01 & 1.42 &1.57 &  90 &  3.52 \\
         &       &    &   76  & 456   &  6 &  7096 &  2.4M &  1.64e-01 & 1.67 &2.00 &  84 &  4.14 \\
         &       &    &   98  & 588   &  6 &  5503 &  1.8M &  1.39e-01 & 1.97 &2.57 &  77 &  4.87 \\
  \hline
  \end{tabular}
  \end{center}

  \begin{center} \begin{tabular}{|c|c|c|c|c|c|c|c|c|c|c|c|c|}
  \hline
  \multicolumn{13}{|c|}{{\bf DNS $5\times 5$, Performance on Mira vs. Summit, $E=8387008$, $N=7$, $n=2.87B$}} \\ 
  \hline
  \hline
  System & Code & Device & Node & Rank & R/N & $E$/R & $n$/Rank &  $t_{step} (s) $ &  R  & R$_{i}$ & eff \% & R$^{**}$ \\
  \hline
  \hline
  Mira    & Nek5000& CPU&    8192 &262144 & 32 &  32   & 10973   &7.00e-01 & 1.00 & 1.00 & 100 & 1.00 \\
  \hline            
  \hline            
          &        &    &   175   & 7350   & 42 &  1141 & 391393  &3.97e+00 & 1.00 & 1.00  & 100 &  0.17 \\
   Summit & Nek5000& CPU&   1152  & 48384  & 42 &  173  & 59456   &9.51e-01 & 4.17 & 6.58  & 63 &   0.73 \\
          &        &    &   2304  & 96768  & 42 &  87   & 29728   &7.30e-01 & 5.43 & 13.16 & 41 &  0.95 \\
  \hline            
  \hline            
          &        &    &  87   &  522   & 6 &  16067&  5.5M &  2.30e-01 & 1.00 & 1.00 & 100 & 3.04  \\  
  Summit  & NekRS  & GPU& 120   &  720   & 6 &  11648&  3.9M &  1.83e-01 & 1.25 & 1.37 & 91  & 3.80  \\
          &        &    & 160   &  960   & 6 &  8736 &  2.9M &  1.49e-01 & 1.53 & 1.83 & 84  & 4.68  \\ 
          &        &    & 220   & 1320   & 6 &  6353 &  2.1M &  1.27e-01 & 1.80 & 2.52 & 71  & 5.48  \\ 
  \hline 
  \end{tabular}
  \end{center}
\caption{\label{mira-table}
Performance on Mira (Nek5000) vs. Summit CPU (Nek5000) and GPU (NekRS).
Timings are in seconds for the wall time per step, $t_{step}$.  R$^{**}$, the
ratio of $t_{step}$ of 8192 nodes on Mira to all others on Summit CPU and GPUs.  
(top) {\bf Spacer-Grid}:  
400 timesteps over simulation time interval [138.0431, 138.0671]
with $\dt$= 6.00e-05 (CFL=1.74) at $Re_D=14000$.  
CHAR=T (1 substep).
(bottom) {\bf DNS $5\times5$}: 
400 timesteps over simulation time interval [59.71, 59.78] with $\dt$= 1.9e-04 (CFL=0.32) at $Re_D=19000$. 
BDF2+EXT2.
}
\end{table*}

\subsection{Summit Scaling Performance}

\begin{figure} 
 \centering
 \includegraphics[width=0.45\textwidth]{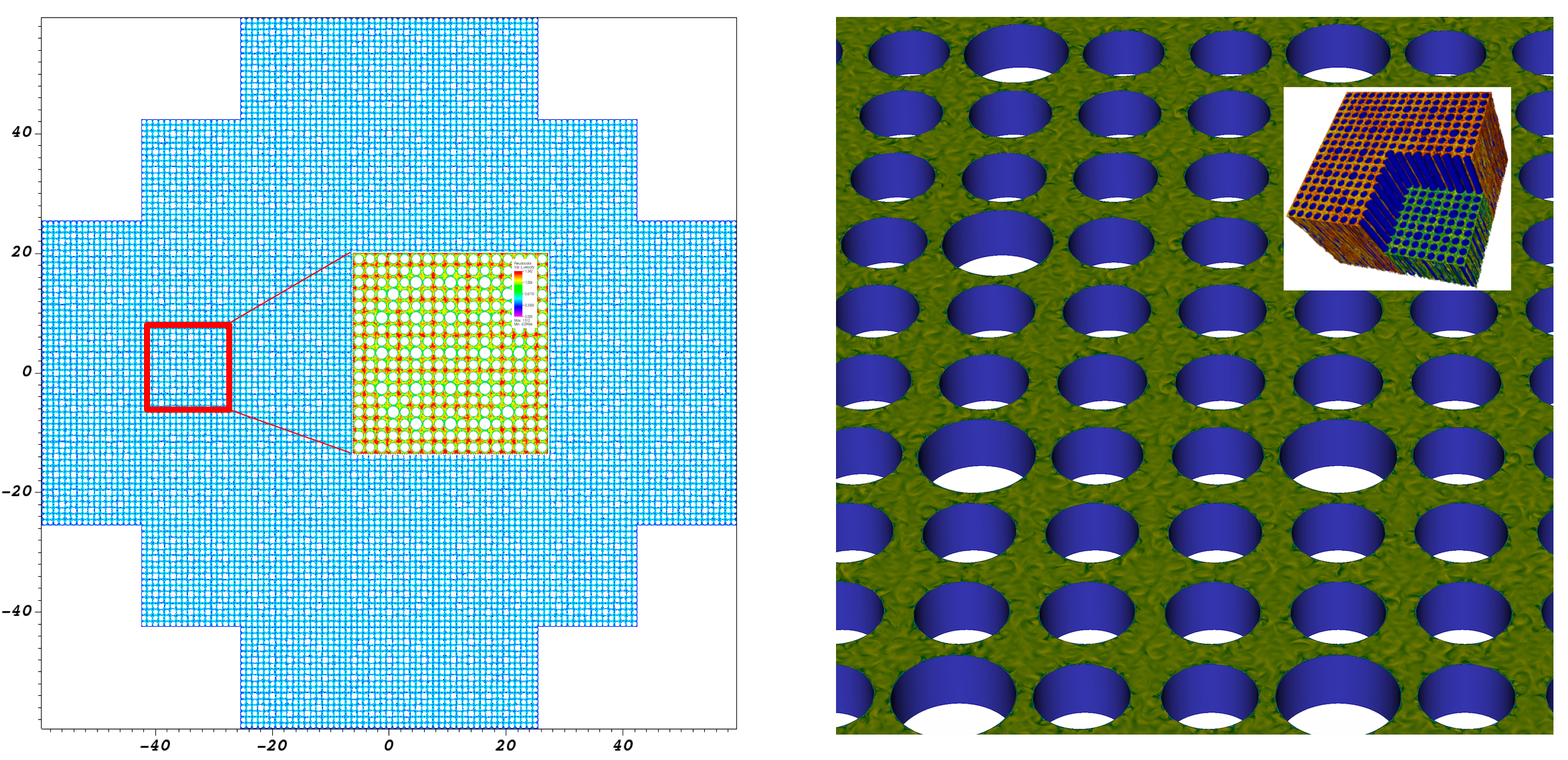}
 \caption{\label{fig:fullcore}
  Full-core and 17$\times$17 rod-bundle configurations.}
\end{figure}

Here we consider simulations scaling out to all of Summit for the rod-bundle
configurations of Fig.~\ref{fig:fullcore}.  Target geometries for small modular
reactors consist of hundreds of 17$\times$17 rod bundles, which total to tens
of thousands of long communicating flow channels.  For scalability tests, we
consider two geometries: a long single 17$\times$17 bundle and a
``full-core'' collection comprising 37 such bundles that are shorter in length.
These cases use inflow-outflow boundary conditions with synthetic vortical
flows as initial conditions.  The pressure iterations are likely to be a bit
higher under fully turbulent conditions, but the overall scaling results for
production runs will be similar to what is presented here.


We measured the average wall time per step in seconds,
$t_{step}$, using 101-200 steps for simulations with $Re_D=5000$. 
The approximation order is $N=7$, and dealiasing is used with $N_q=9$.
We use projection in time, CHEBY+ASM, and flexible PCG for 
the pressure solves with tolerance 1.e-04.  The velocity solves use
Jacobi-PCG with tolerance 1.e-06. BDF3+EXT3 is used for timestepping 
with $\dt$= 3.0e-04, corresponding to 
CFL=0.66 for the full-core case and CFL=0.54 for 17$\times$17 case.
We also show the average velocity ($v_i$) and pressure ($p_i$) iteration
counts over the same simulation interval.
The geometries for the weak-scaling studies were generated by extruding
layers of 2D elements in the axial flow direction.  For strong scaling, we 
used $E=$175M, totaling $n=$60 billion grid points.   

The scaling results are presented in Table~\ref{large-strong-weak}.  
The pressure iteration counts, $p_i \sim$ 2, are lower for these cases
than for the pebble cases, which have $p_i\sim$ 8 for the same timestepper
and preconditioner.  The geometric complexity of the rod bundles is relatively
mild compared to the pebble beds. Moreover, the synthetic initial condition
does not quickly transition to full turbulence.  We expect more pressure
iterations in the rod case (e.g., $p_i \sim$ 4--8) once turbulent 
flow is established.

We observe that these cases exhibit excellent strong scaling to all of Summit,
pointing to $n/P \approx 2.5$M as the 80\% efficiency level for the V100s.
Note that, save for the anomolous DNS 5$\times$5 case, this value of $n/P$ is
consistent with our previous results.  As discussed in
\cite{fischer20a,fischer15}, the leading indicator of parallel scalability
for a given algorithm-architecture coupling is $n/P$, rather than the number of
processing units, $P$.

The weak-scaling results are less straightforward to interpret. 
First, we note that they are conducted at $n/P=$2.1M, which is
beyond the strong-scale limit.  The data is thus heavily influenced
by communication overhead.
Nonetheless, the Rod-1717 case exhibits reasonable weak scaling, with 
only an 18\% drop in efficiency over a 53-fold increase in processor count.
Weak scaling for the full-core case, however, drops to 54\% with only 
a 17-fold increase in processor count.
Part of the performance degradation stems from the very low time exhibited
by the full core for $P=1626$ V100s, which at .066 s is substantially
below the best time of .086 s for the Rod-1717 case.   By contrast, the
weak-scale time for Full-Core is 20\% higher than Rod-1717 for $P=27648$.
Here, the discrepency is closely correlated with the maximum number of 
neighbors that any processor (GPU) is connected to in the $QQ^T$ graph,
which is indicated by ngh in the last column of Table~\ref{large-strong-weak}.
The full-core geometry is a relatively flat graph, and it appears that 
the partitioning for $P=271$ resulted in a 2D decomposition, with a
maximum of 9 neighbors.  When this mesh is partitioned further by
RSB, it results in neighbor counts that are twice those of the Rod-1717 
case.   These results, coupled with the high latency of GPUs (as indicated 
by the large $n_{0.8}$ values), suggests that partitioning with
the aim of minimizing the number of neighbors, rather than the data
volume, might be beneficial in this context.

\begin{table*} 
 \footnotesize
\begin{center} \begin{tabular}{|l|c|c|l|c|c|c|c|c|c|c|c|c|}
  \hline
  \multicolumn{13}{|c|}{{\bf NekRS Strong Scaling on Full Summit, $N=7$, $n=59B$ (full-core) and $n=60B$ (rod-1717)}}\\
  \hline
Case      & Node & GPU &  $E$       & $E$/GPU & $n$/GPU& $v_i$ & $p_i$ & $t_{step} (s) $ &  R  & R$_{\rm ideal}$ & P$_{\rm eff}$  & ngh  \\
  \hline
  \hline
          & 1810 & 10860 & 174233000 & 16044 & 5.5M   &  3 &2  & 2.17e-01 & 1.00 & 1.00  & 100   &  41 \\
          & 2715 & 16290 & 174233000 & 10696 & 3.6M   &  3 &2  & 1.39e-01 & 1.55 & 1.50  & 103   &  26 \\
Full-Core & 3620 & 21720 & 174233000 & 8021  & 2.7M   &  3 &2  & 1.18e-01 & 1.84 & 2.00  & 92    &  32 \\
          & 4525 & 27150 & 174233000 & 6417  & 2.2M   &  3 &2  & 1.22e-01 & 1.76 & 2.50  & 70    &  47 \\
          & 4608 & 27648 & 174233000 & 6301  & 2.1M   &  3 &2  & 1.21e-01 & 1.79 & 2.54  & 70    &  40 \\
 \hline
 \hline
 \hline
         &1810 &10860 & 175618000 & 16171 & 5.5M &3 & 2 & 1.855e-01 &1.00 & 1.00  & 100 & 25 \\
         &2536 &15216 & 175618000 & 11542 & 3.9M &3 & 2 & 1.517e-01 &1.22 & 1.40  &  87 & 25 \\
Rod1717  &3620 &21720 & 175618000 &  8085 & 2.7M &3 & 2 & 1.120e-01 &1.65 & 2.00  &  82 & 26 \\
         &4180 &25080 & 175618000 &  7002 & 2.4M &3 & 2 & 1.128e-01 &1.64 & 2.30  &  71 & 28 \\
         &4608 &27648 & 175618000 &  6351 & 2.1M &3 & 2 & 1.038e-01 &1.78 & 2.54  &  70 & 29 \\
  \hline
 \end{tabular}
\end{center}
\begin{center} \begin{tabular}{|l|c|c|l|c|c|c|c|c|c|c|c|c|}
  \hline
  \multicolumn{13}{|c|}{{\bf NekRS Weak Scaling on Full Summit, $N=7$}}\\
  \hline
Case      & Node & GPU &  $E$       & $E$/GPU & $n$/GPU& $v_i$ & $p_i$ & $t_{step} (s) $ &  R  & R$_{\rm ideal}$ & P$_{\rm eff}$  & ngh  \\
  \hline
  \hline
          & 271  & 1626  & 10249000  & 6303  & 2.1M  &   3 &  2 &  6.58e-02 &   1.00 &   1.00 &   100  &    9\\
          & 813  & 4878  & 30747000  & 6303  & 2.1M  &   3 &  2 &  9.68e-02 &   0.67 &   1.00 &   67.9 &   12\\
Full-Core & 1626 & 9756  & 61494000  & 6303  & 2.1M  &   3 &  2 &  1.05e-01 &   0.62 &   1.00 &   62.5 &   44\\
          & 3253 & 19518 & 122988000 & 6301  & 2.1M  &   3 &  2 &  1.18e-01 &   0.55 &   1.00 &   55.8 &   56\\
          & 4608 & 27648 & 174233000 & 6301  & 2.1M  &   3 &  2 &  1.21e-01 &   0.54 &   1.00 &   54.0 &   40\\
 \hline                                                            
 \hline
 \hline
          &87   & 522   & 3324000    & 6367  & 2.1M  &   3 &  2 &  8.57e-02  &  1.00  &  1.00  &  100   &  25 \\
          &320  & 1920  & 12188000   & 6347  & 2.1M  &   3 &  2 &  8.67e-02  &  0.98  &  1.00  &  98.7  &  25 \\
 Rod-1717 &800  & 4800  & 30470000   & 6347  & 2.1M  &   3 &  2 &  9.11e-02  &  0.94  &  1.00  &  94.0  &  25 \\
          &1600 & 9600  & 60940000   & 6347  & 2.1M  &   3 &  2 &  9.33e-02  &  0.91  &  1.00  &  91.8  &  27 \\
          &3200 & 19200 & 121880000  & 6347  & 2.1M  &   3 &  2 &  9.71e-02  &  0.88  &  1.00  &  88.2  &  25 \\
          &4608 & 27648 & 175618000  & 6351  & 2.1M  &   3 &  2 &  1.03e-01  &  0.82  &  1.00  &  82.5  &  29 \\
 \hline
 \end{tabular}
\end{center}
\caption{\label{large-strong-weak}
Summit strong and weak scalings for full-core and 17$\times$17 rod bundle geometries.
}
\end{table*}

\subsection{Performance on Other GPU Architetures}

Here we constrast our baseline Summit performance with recently deployed
NVIDIA A100 and AMD MI100 node architectures.

The first case is the NekRS turbulent pipe flow example with a synthetic
initial condition, $Re_D= DU/\nu = 19,000$, $E= 6840$, $N=7$, and
$n=n/P=2,346,120$ (which is near the strong-scale performance limit).  We use
characteristics with two RK4 substeps, dealiasing with $N_q=10$, and timestep
size $\dt=.006D/U$, where $U$ is the mean velocity and $D$ is the diameter.
The pressure solve uses projection in time, CHEBY-ASM smoothing for flexible
CG, and a tolerance of 1.e-04.  The Jacobi-PCG tolerance for velocity is
1.e-06.  Timings are in seconds for the averaged-walltime per step, $t_{step}$,
using steps 101--200.  

Table~\ref{pipe-baseline} provides preliminary single-GPU results for AMD 
GPUs against the Summit baseline.  The AMD MI60 and MI100 results were
obtained on the HPE {\em Tulip} platform while the NVIDIA A100 runs were 
done on ALCF's {\em Theta-GPU}.  Performance on a single CPU core and multiple
cores on Summit IBM Power9 is also presented.  The AMD interface is provided by
OCCA's HIP backend.  To produce optimized code, hand-tuning is 
still required for good performance each of the devices.

In Table~\ref{pipe-baseline} we see that Summit is slightly faster than the
Tulip V100, which might be expected given that Summit uses NVLink vs. the PCI-E
interconnect on Tulip.  The NVIDIA A100 clearly is outperforming
the V100 by $1.5\times$, which is in line with the improved memory bandwidth of
the A100.  The (early) MI100 and MI60 GPUs are delivering 85 and 60\% of Summit's
V100, respectively.  We also observe that
Summit's single V100 is comparable to 336 CPU cores on 8 nodes, while
it is only 248$\times$ faster than a single CPU.  The implication is that
the parallel efficiency for NekRS on Summit's Power9 CPUs is 74\% with
$n/P= 6615$. 

\begin{figure} 
 \centering
 \includegraphics[width=0.45\textwidth]{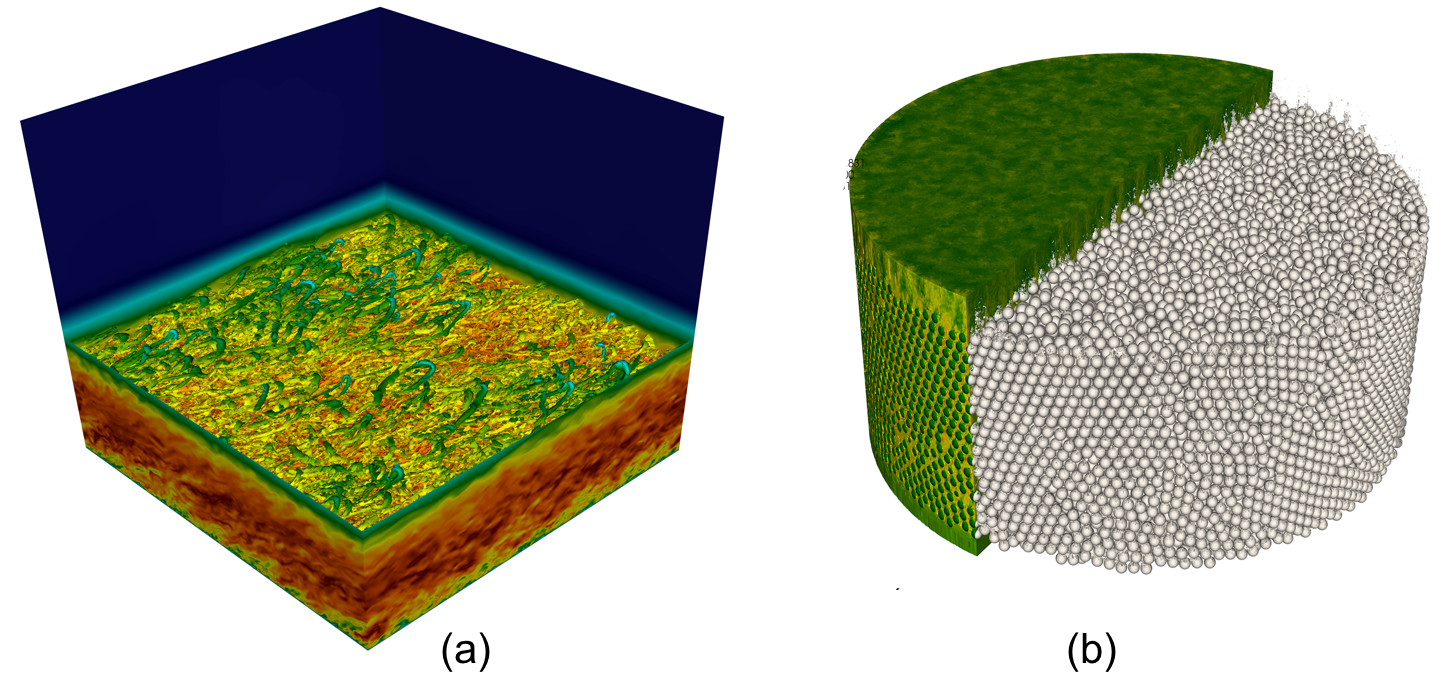}
 \caption{\label{fig:abl-pebble}
  Turbulence in stratified ABL and 44257-pebble configurations.
}
\end{figure}

In our second comparative study we consider multi-GPU, single-node performance
for the Summit V100 vs. ThetaGPU A100s for the atmospheric boundary layer (ABL)
example of Fig.~\ref{fig:abl-pebble} (left).
The domain is doubly-periodic (400m $\times$ 400m $\times$ 400m) with
$E=32768$ spectral elements of order $N=7$ (i.e., $n$=11.2M). 
A geostrophic wind speed of 8 m/s and reference potential temperature of 263.5K
are prescribed with a no-slip condition at the lower wall.  A restart file at
convective time $t$=1710  is used to provide a turbulent initial condition,
corresponding to the physical convective time of 6 hours.  Single-node scaling
shows the 80\% strong-scale limit to be 1.8M points/GPU for both the V100 and
A100, with the A100 running at .055 s/step and 1.55 times faster than the V100.
We remark that the low strong-scale limit of $n/P=1.8$M for these single-node 
studies is more likely due to the low number of neighbors (at most 6 or 8
for the V100 or A100, respectively), rather than a high internode communication
cost in the multinode cases.

We close with a final example illustrating the potential of GPU-based
simulations of turbulence in HPC settings.  The example, shown 
on the right in Fig.~\ref{fig:abl-pebble},
is a 44,257-pebble configuration,
which is a 
prototype for pebble-bed reactors that will ultimately hold hundreds
of thousands of spherical pebbles.  This example has 13M elements of
order $N=7$ ($n=4.5$B).  
The all-hex mesh was developed from an initial Voronoi tessellation of the
sphere centers, with each Voronoi facet tessellated into quadrilaterals
that are then projected onto the sphere surfaces in order to sweep out
a hexadral volume.  Edge collapse, mesh refinement, and mesh smoothing
tools ensure a high-quality mesh for the fluid flow in the void space.
Timestepping is based on 2nd-order characteristics with a single RK4 subcycle,
$N_q=11$, and $\dt$ =3.e-4 (CFL=4).   The average number of velocity
iterations (tol=1.e-6) is 3, and the average number of pressure iterations
is 18 (tol=1.e-4).   On 1,788 V100s ($n/P=2.5$M), $t_{step}$=.54 s.
The timing breakdown is 
10\% for advection (\ref{eq:hyp}), 6\% for the velocity solve (\ref{eq:ns_2d}),
and 84\% for the pressure solve.
The pressure solve is broken down into percentage of total simulation time:
16\% for the coarse-grid solve and
56\% for the remainder of the preconditioner.
While the time per step is higher than for the other cases, these
simulations strong-scale well; and the target configuration 
of 300,000 pebbles, which will require about 30B grid points, 
is well within the current performance envelope on Summit.


\begin{table*} 
  \footnotesize
  \begin{center} \begin{tabular}{|l|l|l|r|r|c|c|c|c|c|}
  \hline
  \multicolumn{10}{|c|}{{\bf NekRS single GPU performance for turbulent pipe flow simulation, $n=2,222,640$}}\\
  \hline
   system & device & API & rank & node &  $E$ & $N$ & $n$/rank & $t_{step} (s) $ &  R \\
  \hline
       OLCF Summit        &  NVIDIA V100 & CUDA &  1  & 1  & 6480 & 7 & 2.22M &  8.51e-02 & 1.00e+00 \\
       ALCF Theta-GPU     &  NVIDIA A100 & CUDA &  1  & 1  & 6480 & 7 & 2.22M &  5.59e-02 & 1.52e+00 \\
       HPE  Tulip         &  NVIDIA V100 & CUDA &  1  & 1  & 6480 & 7 & 2.22M &  8.85e-02 & 9.61e-01 \\
       HPE  Tulip         &  AMD MI100   & HIP  &  1  & 1  & 6480 & 7 & 2.22M &  9.96e-02 & 8.54e-01 \\
       HPE  Tulip         &  AMD MI60    & HIP  &  1  & 1  & 6480 & 7 & 2.22M &  1.41e-01 & 6.03e-01 \\
       OLCF Summit        &  IBM Power9  &  C   &  1  & 1  & 6480 & 7 & 2.22M &  1.99e+01 & 4.27e-03 \\
       OLCF Summit        &  IBM Power9  &  C   & 336 & 8  & 6480 & 7 & 6615  &  8.02e-02 & 1.06e+00 \\
      \hline
  \end{tabular}
\end{center}
\caption{\label{pipe-baseline}
NekRS baseline performance on a single GPU of HPE Tulip AMD Instinct\textsuperscript{TM} MI100, 
AMD Radeon Instinct\textsuperscript{TM} MI60, 
and Nvidia V100 PCle and ALCF/Theta-GPU Nvidia A100 SXM2, compared to  OLCF/Summit Nvidia V100 SXM2, 
for turbulent pipe flow simulation with $Re= 19,000$, $E= 6840$, and $N=7$.}
\end{table*}

 \begin{table*} [t]
  \footnotesize
  \begin{center} \begin{tabular}{|c|c|c||c|c|c||c|c|c||c||c|}
  \hline
  \multicolumn{11}{|c|}{{\bf NekRS on a Singe Node for Atmospheric Boundary Layer Model, $E=32768$, $N=7$, $n=11,239,424$}} \\
  \hline
         gpu   & E/gpu& n/gpu & V100  & R & P$_{\rm eff}(\%)$ & A100  & R  &P$_{\rm eff}(\%)$ & R$_{\rm ideal}$ & R$_{V/A}$\\
  \hline                                                                                            
           2   & 16384 & 5.6M  & 2.050e-01  & 1.00  &  100     & 1.341e-01   & 1.00 & 100   & 1.00 & 1.52      \\
           3   & 10923 & 3.7M  & 1.464e-01  & 1.40  &  93.3    & 9.544e-02   & 1.40 & 93.7  & 1.50 & 1.53      \\
           4   & 8192  & 2.8M  & 1.171e-01  & 1.75  &  87.5    & 7.485e-02   & 1.79 & 89.6  & 2.00 & 1.56      \\
           5   & 6553  & 2.2M  & 9.898e-02  & 2.07  &  82.8    & 6.371e-02   & 2.10 & 84.2  & 2.50 & 1.55      \\
           6   & 5461  & 1.8M  & 8.575e-02  & 2.39  &  79.7    & 5.519e-02   & 2.43 & 81.0  & 3.00 & 1.55      \\
           7   & 4681  & 1.6M  &     -      &   -   &    -     & 5.080e-02   & 2.64 & 75.4  & 3.50 &   -       \\
           8   & 4096  & 1.4M  &     -      &   -   &    -     & 4.545e-02   & 2.95 & 73.8  & 4.00 &   -       \\
  \hline
  \end{tabular}
\end{center}
\caption{\label{nek-abl-node1} Atmospheric boundary layer baseline performance on Summit V100 SXM2 and ThetaGPU A100 SXM4.}
\end{table*}
\section{Conclusions}
We developed an C++/OCCA-based open-source Navier--Stokes solver for GPUs 
that leverages prior scaling development in Nek5000 and high-performance
kernels developed in libParanumal.  We discuss its performance and scalability
on leadership computing platforms.  The solver is based on 2nd-
or 3rd-order timesplitting of the incompressible Navier--Stokes equations with
an exponentially convergent spectral-element-based discretization in space.
We demonstrate weak- and strong-scaling up
to 27,648 V100 GPUs on OLCF's Summit system for 
reactor geometries with problem sizes of more than 175M spectral
elements ($n=60$B gridpoints).  Performance results show that NekRS
sustains 80--90\% of the realizable (bandwidth-limited) peak and that
80\% parallel efficiency on Summit is realized for local problem sizes
of $n/P \approx 2.5$M, where $P$ is the number of GPUs employed for the
simulation.  Preliminary timing data for NVIDIA A100s and AMD MI100s are
also presented.


\section*{Acknowledgments}

This material is based upon work supported by the U.S. Department of Energy, 
Office of Science, under contract DE-AC02-06CH11357.

This research is supported by the Exascale Computing Project (17-SC-20-SC), a
collaborative effort of two U.S. Department of Energy organizations (Office of
Science and the National Nuclear Security Administration) responsible for the
planning and preparation of a capable exascale ecosystem, including software,
applications, hardware, advanced system engineering and early testbed platforms,
in support of the nation's exascale computing imperative.

The research used resources of the Argonne Leadership Computing Facility, which
is supported by the U.S. Department of Energy, Office of Science, under Contract
DE-AC02-06CH11357. This research also used resources of the Oak Ridge Leadership
Computing Facility at Oak Ridge National Laboratory, which is supported by the
Office of Science of the U.S. Department of Energy under Contract DE-AC05-00OR22725.
Support was also given by the Frontier Center of Excellence. 



\bibliographystyle{bst/elsarticle-num-names}
\bibliography{scale}

\newpage
The submitted manuscript has been created by UChicago Argonne, LLC, Operator of 
Argonne National Laboratory (``Argonne"). Argonne, a U.S. Department
of Energy Office of Science laboratory, is operated under Contract No. DE-AC02-06CH11357. 
The U.S. Government retains for itself, and others acting on its behalf,
a paid-up nonexclusive, irrevocable worldwide license in said article to reproduce,
prepare derivative works, distribute copies to the public, and perform publicly and
display publicly, by or on behalf of the Government.
The Department of Energy will provide public access to these results of federally sponsored research in accordance with the DOE Public Access Plan. http://energy.gov/downloads/doe-public-access-plan.

ECP Disclaimer: This research is supported by the Exascale Computing Project
(17-SC-20-SC), a collaborative effort of two U.S. Department of Energy
organizations (Office of Science and the National Nuclear Security
Administration) responsible for the planning and preparation of a capable
exascale ecosystem, including software, applications, hardware, advanced system
engineering and early testbed platforms, in support of the nation’s exascale
computing imperative.

\end{document}